\documentclass{ws-p8-50x6-00}

\begin{document}

\title{
Confinement in the Big Bang and Deconfinement in the Little Bangs
at CERN-SPS\footnotemark}\footnotetext{$^*$ Work supported by 
BMBF 06DR921, WTZ UKR-008-98 and
STCU-015.}

\author{B. K\"ampfer, K. Gallmeister and  O.P. Pavlenko}

\address{Forschungszentrum Rossendorf, PF 510119, 01314 Dresden, 
Germany\\
E-mail: kaempfer@fz-rossendorf.de}


\maketitle

\abstracts{The evolution of strongly interacting matter during the
cosmological confinement transition is reviewed. Despite of many 
proposed relics no specific signal from the rearrangement of quarks
and gluons into hadrons has been identified by observations.
In contrast to this, several observables in heavy-ion collisions
at CERN-SPS energies point to the creation of a matter state
near or slightly above deconfinement. We focus here on the analysis
of dileptons and direct photons.
Similarities and differences of the Big Bang and the Little Bang
confinement dynamics are elaborated.}

\section{Introduction}

The theory of strong interaction, QCD, points to a transition
from a confined hadronic phase to the quark-gluon plasma at
sufficiently high temperatures. The deconfinement temperature, 
$T_c$, is in the order of 170 MeV or slightly
larger.\cite{Karsch_1} The very nature of the deconfinement 
transition depends on
yet poorly constrained parameters, such as quark masses (cf.\ 
Ref.\cite{Karsch_2}). Above $T_c$ the recent advanced QCD lattice
calculations \cite{Karsch_3} deliver results on the equation of state 
of partonic matter, which can be understood within quasi-particle
models.\cite{Peshier} Fig.~1 shows a few examples.

\begin{figure}[t]
 \vskip -.01cm
 \epsfxsize=9.1pc \epsfbox{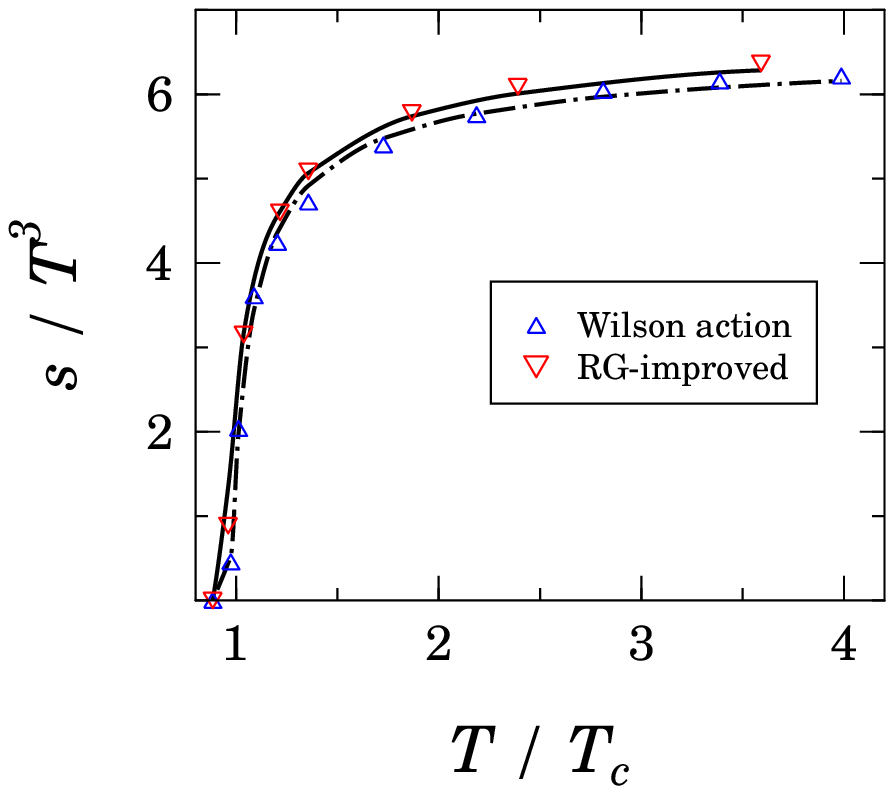} \hfill
 \epsfxsize=9.1pc \epsfbox{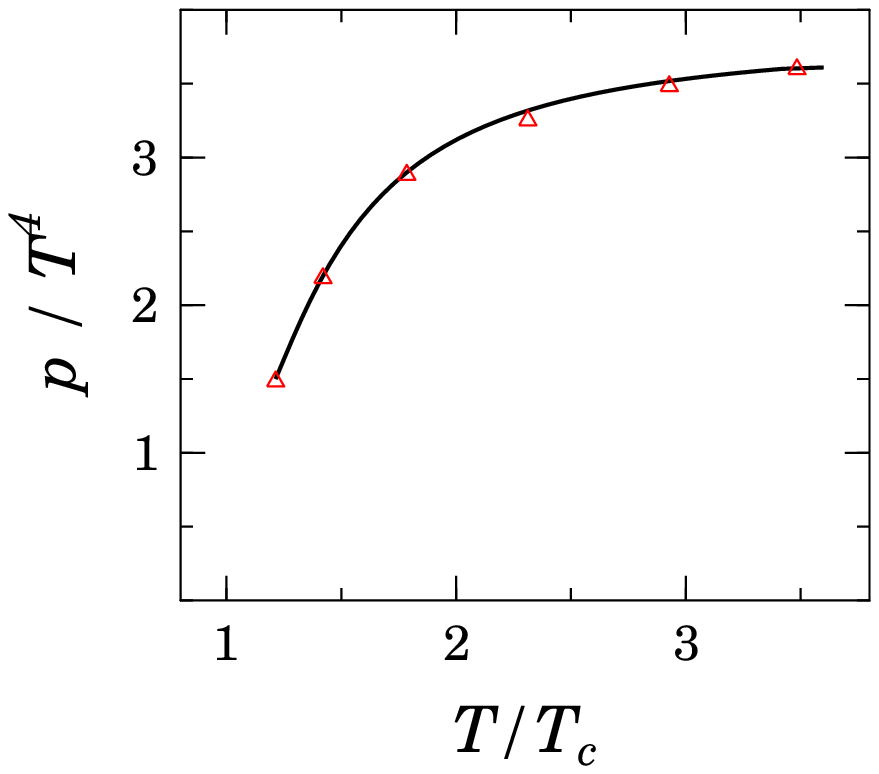} \hfill
 \epsfxsize=9.1pc \epsfbox{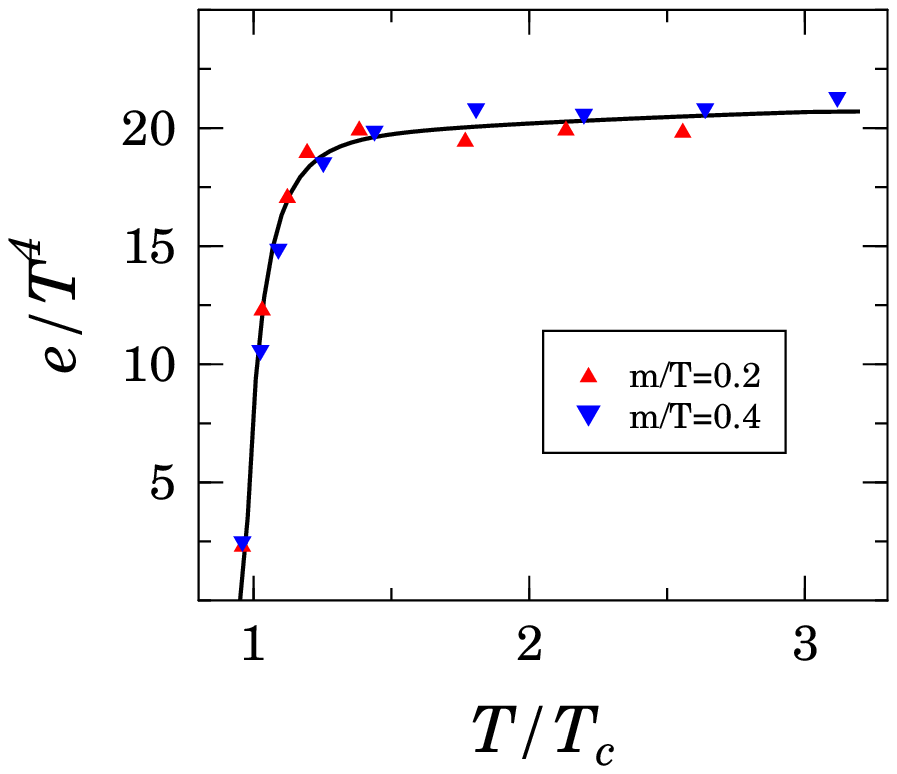}
 \caption{Equation of state of a gluon gas (left panel, entropy
 density $s$) and
 a two-flavor (middle panel, pressure $p$) and a four-flavor
 (right panel, energy density $e$) quark-gluon plasma
 as a function of the scaled temperature. Lattice QCD results
 (symbols) from the Bielefeld group; the curves represent an adjusted
 quasi-particle model (for details consult 
 Ref.\protect\cite{Peshier}).}
 \label{lattice_QCD}
\end{figure}

Hot deconfined matter must have existed in the
early universe. According to the standard Big Bang cosmology the
thermalized matter in the universe undergoes continuous cooling.
That means that in the Big Bang
the confinement transition at $T_c$ happened,
where quarks and gluons become strongly correlated, thus forming
particles with large masses, such as protons and neutrons,
and other hadrons as well. One intriguing question is whether the
cosmic confinement transition left some specific imprint on the
subsequent evolution of matter or a verifiable direct signal.
Despite of many proposed relics, up to now no specific signal has
been found. 

One of the primary goals of the
investigations of heavy-ion collisions at the CERN-SPS is the hunt for
signals from deconfined matter. Indeed, there are
indications that the quark-gluon plasma has already been
encountered.\cite{CERN} In Sec.\ \ref{dileptons} we shall consider
in some detail the electromagnetic radiation from the hot fireballs
created in the collisions and deduce a temperature scale of
${\cal O}(T_c)$ from the data. Therefore, one can conclude that
in the Little Bangs at CERN-SPS a matter state is created being
near the borderline of confined and deconfined matter.
This matter state resembles, to some extent as explained below,
the matter in the universe at temperatures around $T_c$.
The starting heavy-ion programme at RHIC and the future experiments
at LHC are aimed at achieving matter states with temperatures
clearly above $T_c$. 

We are going to
elaborate the similarities and differences of the confinement transition
in the Big Bang and the Little Bangs (Sec.\ 2). In particular, we
review possible relics of the cosmological confinement transition
(Sec.\ 3).
Then we present an analysis of dilepton and photon spectra observed
in Little Bangs (Sec.\ 4).
The conclusions can be found in Sec.\ 5.

\section{Big Bang versus Little Bang dynamics}

As starting point of describing the evolution of matter
we chose relativistic hydrodynamics which is standard in cosmology
\cite{Coles_Lucchin}
and which has been proven to be useful for heavy-ion 
collisions.\cite{Greiner_Stocker}
The dynamics of matter is governed by the local energy-momentum
conservation, 
\begin{equation}
T^{ij}{}_{;j} = 0, 
\end{equation}
where we approximate the energy-momentum tensor by that of a 
perfect fluid,
$
T_{ij} = e u_i u_j + p (u_i u_j - g_{ij})
$
with four-velocity $u^i$ obeying $u^i u_i = + 1$, 
$g_{ij}$ is the metric tensor,
$e$ stands for the total energy density,
and $p$ denotes the thermodynamic pressure;
the semicolon denotes the covariant derivative.

\subsection{Friedmann's equation}
To cast Eq.\ (1) into a tractable form one has to specify the
space-time symmetry and the flow pattern. Basing on the cosmological
principle\footnote{This states homogeneity and isotropy in the 3D
configuration space, which seem to be proven at early times by the tiny
temperature fluctuations of $\Delta T / T < 10^{-4}$ of the 
present background 
radiation emerged from photon freeze-out at a world age of 300,000
years at temperature of 3,000 K.}
and Einstein's field equations for geometrodynamics one gets
Friedmann's equation
\begin{equation}
\dot e = - \frac{3}{M_{\rm Pl}} 
\sqrt{\frac{8 \pi}{3}} (e + p) \sqrt{e}
\label{Friedmann}
\end{equation}
for the time evolution of the total energy density of matter;
here $M_{\rm Pl}$ is the Plank mass determining even nowadays the 
cosmic dynamics.

The recent discovery of an accelerated expansion of the universe
points to a substantial contribution of either a vacuum energy
density and pressure, $e^{\rm vac}$ and $p^{\rm vac} = - e^{\rm vac}$,
or a {\em quintessence} which dynamical behavior is not yet settled.
Also the back extrapolation of the dark matter contribution meets
uncertainties. With these caveats in mind we include in $e$ and $p$
only thermal excitations. The dynamical time scale is, from
Eq.\ (\ref{Friedmann}),
\begin{equation}
\frac{e + p}{\dot e} \sim \frac{M_{\rm Pl}}{\sqrt{e}}
\sim 10^{19} \, \mbox{fm/c \, at\, } \, T \sim 200 \, \mbox{MeV},
\end{equation}
showing that large energy densities drive fast evolution.
This time scale is so large that quarks and gluons or, later on,
hadrons
and all leptons and the photons are in thermal and chemical 
equilibrium. These equilibrium conditions, however, cause
a memory loss and, as we shall see below, little chances to find
specific relics unless such ones which drop out of equilibrium.

\subsection{Bjorken's equation}

At sufficiently high energies, parton cascade and string models 
for describing heavy-ion collisions point to a dominant longitudinal
motion of matter with four-velocity
$u^i = \gamma (1,0,0,v_z)$ with $v_z = z/t$ and
$\gamma = (1 - v_z^2)^{-1/2}$.
With this flow pattern, Eq.\ (1) becomes
the celebrated Bjorken equation
\begin{equation}
\dot e = - \frac{1}{\tau} (e + p)
\label{e_dot}
\end{equation}
being the Little Bang pendant to the Friedmann's Eq.\ 
(\ref{Friedmann}) for Big Bang.
Thermalization sets in at $\tau_1 = {\cal O}$(1 fm/c), therefore the
dynamical time scale is
\begin{equation}
\frac{e + p}{\dot e} = \tau \sim \tau_1
\sim 1 \, \mbox{fm/c}.
\end{equation}  
Due to the shortness of this scale and the smallness of the considered
systems, photons and leptons, once created, cannot come to equilibrium
with the strongly interacting matter, rather they leave the fireballs
nearly undisturbed and carry information on
the early hot stages, where only strongly interacting matter
can achieve local thermal equilibrium.
The chemical equilibrium can also terminate early, thus opening
another window to primordial stages.

Before discussing signals from hot matter in heavy-ion
collisions let us consider possible relics from the confinement
transition in Big Bang.

\section{Relics of the cosmic confinement transition?}

A discussion of this topic is hampered by the mentioned uncertainties
of the nature of the deconfinement transition and the behavior of
confined matter near $T_c$. One can use, for instance, various 
condensates, in particular the chiral condensate, as order parameters
characterizing confinement. Within our phenomenological approach
one has to resort to the behavior of the equation of state which may
display a first-order phase transition (for light quarks) or a sharp
cross over (for light u, d quarks and medium-heavy s quarks).
Supposed ones describes with a bag model equation of state
the deconfined matter, $p = \frac 13 (e - 4 B)$ (here $B$ parameterizes
the vacuum energy density), and with $p = \frac 13 e$ the hadronic
matter and adds appropriately the background contribution of photons and
leptons, one finds the 
beginning of the transition at world age $t_1 \sim 6$ $\mu$sec and the
end of the transition in case of a near to equilibrium transition
with small surface tension $t_2 \sim 12$ $\mu$sec.\footnote{General 
characteristic quantities at $T_c$ are:
horizon radius $R_H \sim 10$ km,
Hubble time $t_H \sim 10^{-5}$ sec,
energy density within the horizon $M_H$ corresponding to 
$\sim 1 M_\odot$, and
baryon charge within horizon $N_H^B \sim 10^{50}$.}

\begin{figure}
 \vskip .01cm
 ~\center
 \epsfxsize=22pc \epsfbox{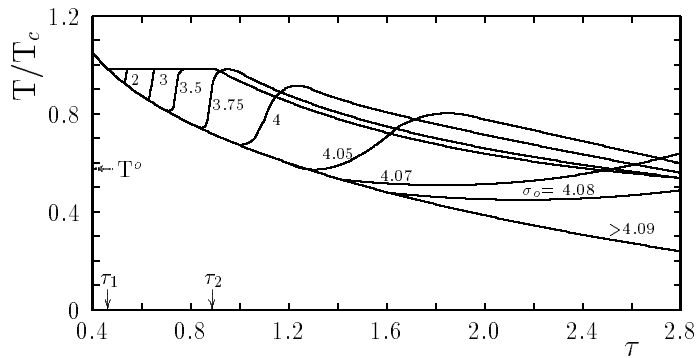}
 \caption{The temperature evolution during a first-order phase transition
 for various values of the surface tension parameter $\sigma_0$
 as a function of the scaled dimensionless time
 $\tau = 2 {\cal C} B^{1/2} t$.
 $\tau_1$ and $\tau_2$ denote the beginning and end of an equilibrium
 transition; $T^o$ is the temperature of the maximum nucleation rate
 (cf.\ Ref.\protect\cite{KLP}),
 and ${\cal C} = \sqrt{8 \pi/3} / M_{\rm Pl}$.}
 \label{T(t)}
\end{figure}

Examples of possible cooling curves within the framework of classical
nucleation theory are displayed in Fig.~\ref{T(t)}
for various values of the surface tension (see Ref.\cite{KLP} for
details). Since lattice QCD calculations point to small values of the
surface tension at the boundary of confined and deconfined matter,
a small supercooling is to be expected. Then frequent bubble
nucleation sets in suddenly after some supercooling,
and the resulting released latent heat
reheates the matter to $T_c$, where nucleation ceases. Bubble growth
determines the further evolution. This bubble growth is accompanied
by shock waves \cite{Gyulassy_McLerran} causing an irregular pattern
of intersecting shock waves. It has been speculated that at these
intersections matter is compressed and seeds for density enhancements
are created. At sufficient density enhancement, matter can collapse
to black holes. 

Within the scenario of a first-order cosmic confinement transition
various other possible imprints have been studied.
Among them are:\\
-- Isothermal baryon fluctuations:
At given temperature the tiny, but nevertheless finite, baryo-chemical
potential gives rise to different baryon densities in the confined
and deconfined phases. If the baryon concentration in the region of 
shrinking quark matter at the end of the confinement era is not
diffused away, then the nucleo-synthesis can be 
affected.\cite{nucleosynthesis}\\
-- Vanishing sound velocity: 
The restoring pressure gradient in density-enhanced regions, created by
fluctuations, vanishes and these regions can collapse in free fall.
On scales being much larger than the typical bubbles, kinetically
decoupled cold dark matter can be trapped in gravitational wells.
Therefore, dark matter candidates can be distributed very
inhomogeneously.\cite{Schwarz}
Also previously causally non-connected weak
fluctuations on super-horizon scales can collapse to 
black holes.\cite{Jedamzik}\\
-- Strangelets and quark nuggets:
Weak processes establish equilibrium in the
reactions $d \leftrightarrow u + l^- + \nu_l$
and
$s \leftrightarrow u + l^- + \nu_l$
and the corresponding
cross channels ($l$ stands here for the electron or the muon).
Thus, a substantial fraction of deconfined matter resides 
strange quarks. It has been speculated that strange
quarks can stabilize quark matter.\cite{strangelets} 
Relics from the 
confinement transition can accordingly exist as stable quark nuggets
or strangelets. The dedicated search for such exotic matter states in
heavy-ion collisions, however, turned out negative.

For a recent review on such and further possible relics 
see Ref.\cite{Annalen}

\section{Analysis of dilepton and photon spectra \label{dileptons}}

By now a wealth of electromagnetic signals from the fireball in the
Little Bangs at CERN-SPS has been registered. The spectra from the
following collaborations are available:
(i) CERES: Pb(158 AGeV) + Au $\to$ $e^+ e^-$,
(ii) NA50: Pb(158 AGeV) + Pb $\to$ $\mu^+ \mu^-$,
(iii) WA98: Pb(158 AGeV) + Pb $\to$ $\gamma$,
(iv) CERES: S(200 AGeV) + Au $\to$ $e^+ e^-$,
(v) NA38: S(200 AGeV) + U $\to$ $\mu^+ \mu^-$,
(vi) HELIOS/3: S(200 AGeV) + W $\to$ $\mu^+ \mu^-$,
(vii) WA80: S(200 AGeV) + Au $\to$ $\gamma$ (only upper bounds).
In a schematic picture the electromagnetic signals can be considered
as superposition of the following sources:
(i) On very short time scales there are hard initial processes
among the partons, being distributed according to primary nuclear
parton distributions, such as the Drell-Yan process
and charm production.
(ii) On intermediate time scales there are the so-called secondary
interactions among the constituents of the hot and dense,
strongly interacting matter. This stage is often denoted as thermal era and
the emitted dileptons as thermal dileptons. 
(iii) If the interactions among the hadrons in a late stage cease, there
are hadronic decays into dileptons and other decay products.

\begin{figure}[t]
 \vskip -.01cm
 \epsfxsize=9pc \epsfbox{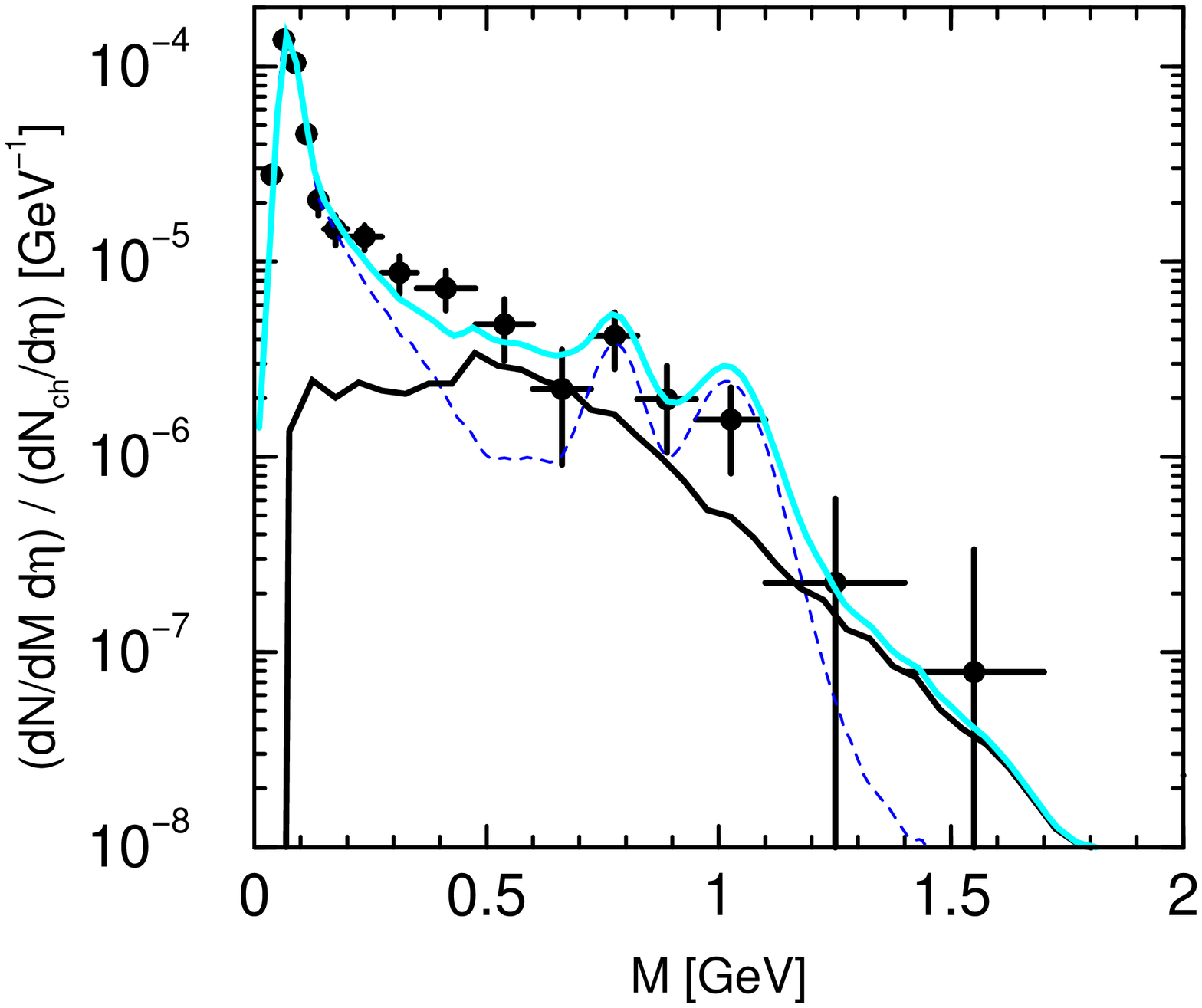} \hfill
 \epsfxsize=8.8pc \epsfbox{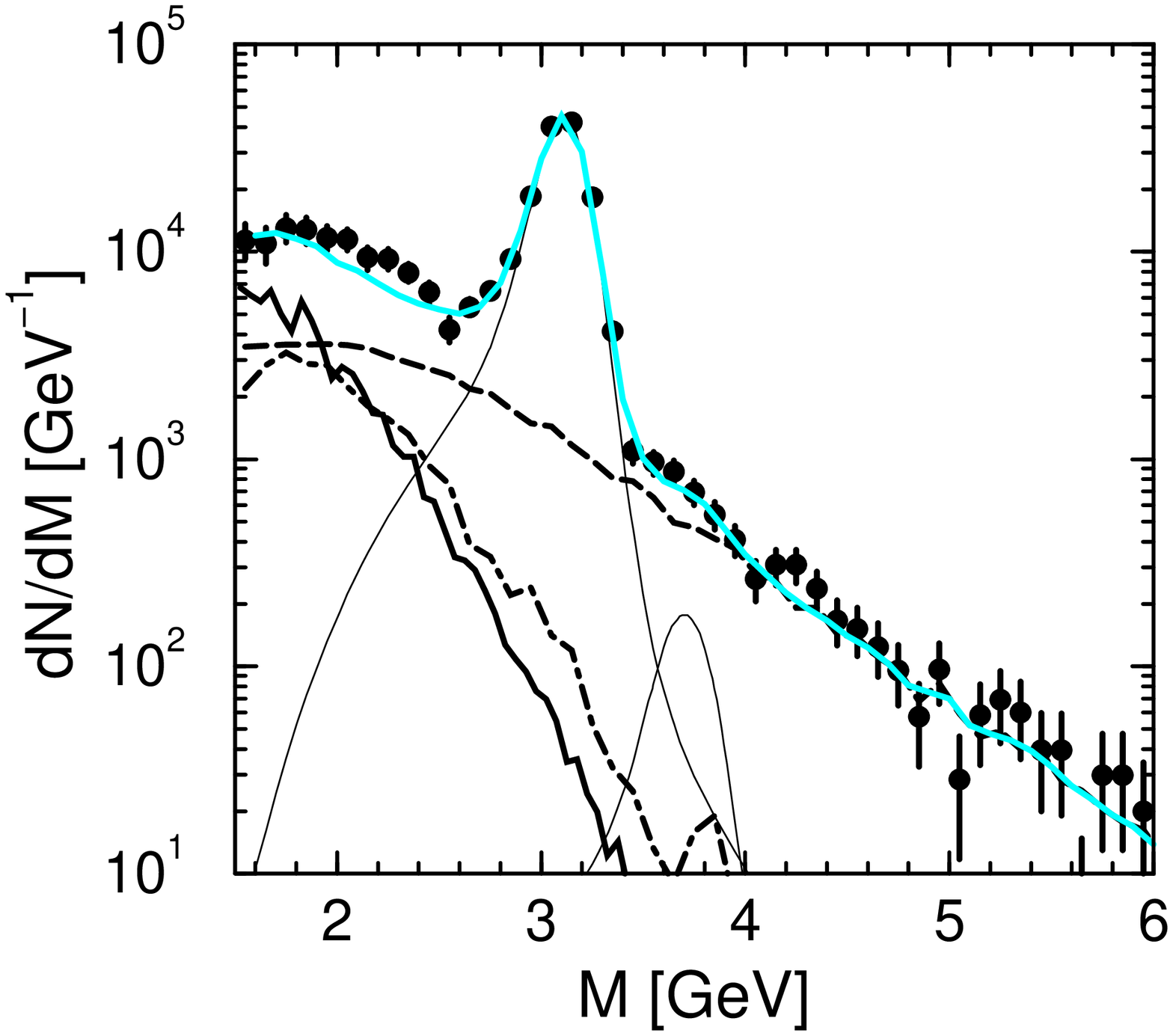} \hfill
 \epsfxsize=9pc \epsfbox{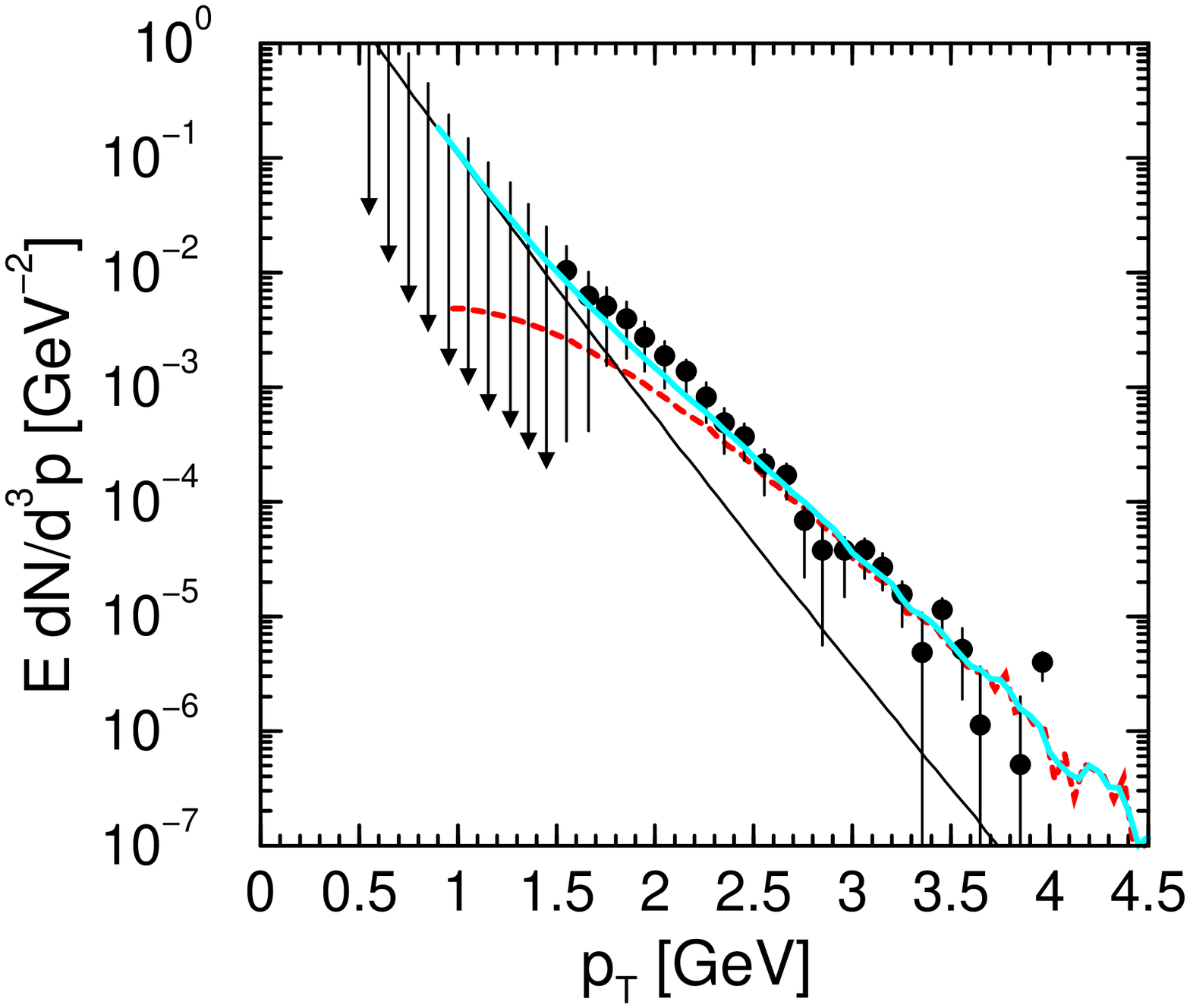}
 \caption{Comparison of our model with dilepton data 
(left panel for the CERES data,\protect\cite{CERES}
 dashed line: hadronic cocktail;
 middle panel for the NA50 data,\protect\cite{NA50}
 dashed line: Drell-Yan contribution,
 dot-dashed line: open charm contribution,
 thin lines: parameterizations of the $J/\psi$ and $\psi'$)
 and
 the photon data (right panel for WA98 data,\protect\cite{WA98}
 dashed line: hard direct photons, $v_0 = 0.3$).
 The thermal contribution (solid curves)
 is characterized by $T_{\rm eff} = 170$ MeV
 and $N_{\rm eff} = 3.3 \times 10^4$ fm${}^4$.
 The sum of all contributions is depicted by the gray curves.}
 \label{em_signals}
\end{figure}

We describe the hard processes by up-scaling the results of the event
generator PYTHIA for pp collisions at appropriate energies (for
details consult Ref.\cite{Gale}).
First we attempt a unifying description of the data by
super-positioning the background (hadronic cocktail, Drell-Yan,
correlated semileptonic decays of open charm-mesons, hard direct
photons) and the thermal source parameterized by
\begin{eqnarray}
\frac{dN_{l \bar l}}{d^4 Q} & = &
\frac{5 \alpha^2}{36 \pi^4} N_{\rm eff}
\exp \left\{ - 
\frac{M_\perp \cosh (Y - Y_{\rm cms} )}{T_{\rm eff}} \right\},
\label{dilepton_parametrization} \\
E \frac{dN_\gamma}{d^3 p} & = &
N_{\rm eff} \, \frac{5 \alpha \alpha_s T_{\rm eff}^2}{12 \pi^2}
\int_0^1 ds \, s^2
\int_{-1}^{+1} d \xi \mbox{e}^{-A} \log
\left[1 + \frac{\kappa}{\alpha_s} A \right],
\label{photon_parametrization}
\end{eqnarray}
where $A = \frac{p_\perp \cosh y \, (1 - s v_0 \xi)}
{T_{\rm eff} \sqrt{1 - (sv_0)^2}}$,
$\kappa = 2.912/(4 \pi )$;
$Q = (M_\perp \cosh Y, M_\perp \sinh Y, \vec Q_\perp)$
and
$p = (p_\perp \cosh y, p_\perp \sinh y, \vec p_\perp)$
are the four-momenta of the the dileptons and photons with
transverse mass $M_\perp = \sqrt{M^2 + Q_\perp^2}$
(here $M$ is the invariant mass), transverse momenta
$Q_\perp$ and $p_\perp$, and rapidities $Y$ and $y$, respectively.
The two parameters
$T_{\rm eff}$ and $N_{\rm eff}$ are to be adjusted to the experimental data.
In Eqs.\ (\ref{dilepton_parametrization},
\ref{photon_parametrization})
the time evolution of the volume of the fireball and the temperature
have been replaced by averages.\cite{Gale,dileptons}
Fig.\ \ref{em_signals} displays a few examples of the quality of our data
description. Other examples, such as the transverse momentum spectra
and an analysis of the sulfur beam induced reactions, can been found
in Refs.\cite{Gale,dileptons} 

The unique outcome of these studies is the value $T_{\rm eff} =$ 160 
$\cdots$ 170 MeV. Being aware of the schematic character of
Eqs.\ (\ref{dilepton_parametrization}, \ref{photon_parametrization})
one can implement a dynamical scenario. With parameters, partially
fixed by hadronic observables, one finds a maximum temperature
${\cal O}(200)$ MeV or slightly above.\cite{dileptons,Alam}
This is the most stringent
proof that at CERN-SPS in heavy-ion collisions such temperatures
are achieved which are in the deconfinement region. Of course,
the effect of a finite baryon density in 
Eqs.\ (\ref{dilepton_parametrization}, \ref{photon_parametrization})
and the assumption of thermalization need further consideration.

\section{Summary}

The analysis of electromagnetic signals emitted in the course
of central heavy-ion collisions at CERN-SPS point to a state 
of strongly interacting matter with temperatures met also at
confinement during the cosmic evolution.
However, the dynamical time scales are vastly different. 
Equilibrium conditions mean memory loss, therefore,
a specific imprint of the cosmic confinement transition has not yet
been identified and seems unlikely.


\begin{thebibliography}{99}
\bibitem{Karsch_1}
F. Karsch, {\it Nucl. Phys. Proc. Suppl.} {\bf 83 - 84}, 14 (2000).
\bibitem{Karsch_2}
A. Peikert {\it et al},
{\it Nucl. Phys. Proc. Suppl.} {\bf 73}, 468 (1999).
\bibitem{Karsch_3}
F. Karsch {\it et al},
hep-lat/0010027,\\
F. Karsch, E. Laermann, A. Peikert, {\it Phys. Lett.} B {\bf 478}, 447
(2000).
\bibitem{Peshier}
A. Peshier, B. K\"ampfer, G. Soff, {\it Phys. Rev.} C {\bf 61}, 045203
(2000),\\
A. Peshier, B. K\"ampfer, O.P. Pavlenko, G. Soff, 
{\it Phys. Rev.} D {\bf 54}, 2399 (1996).
\bibitem{CERN}
U. Heinz, hep-ph/0009170, {\it Nucl. Phys.} A in print.
\bibitem{Coles_Lucchin} P. Coles, F. Lucchin, {\it Cosmology},
(John Wiley \& Sons, Chichester New York Brisbane Toronto Singapore 1995).
\bibitem{Greiner_Stocker} W. Greiner, H. St\"ocker, 
{\it Phys. Rep.} {\bf 137}, 277 (1986). 
\bibitem{KLP} B. K\"ampfer, B. Luk\'acs, G. Pa\'al,
{\it Cosmic phase transitions} (Teubner-Verlag, Stuttgart Leipzig 1994).
\bibitem{Gyulassy_McLerran} 
J. Ignatius {\it et al}, {\it Phys. Rev.} D {\bf 49}, 3854 (1994), 
{\bf 50}, 3738 (1994),\\
M. Gyulassy {\it et al}, {\it Nucl. Phys.} B {\bf 237}, 477 (1984).
\bibitem{nucleosynthesis}
M.B. Christiansen, J. Madsen, {\it Phys. Rev.} D {\bf 53}, 5446 (1996),\\
I.S. Suh, G.J. Mathews, {\it Phys. Rev.} D {\bf 58}, 123002 (1998).
\bibitem{Schwarz}
C. Schmid, D.J. Schwarz, P. Widerin, {\it Phys. Rev.} 
D {\bf 59}, 043517 (1999),\\
D.J. Schwarz, {\it Mod. Phys. Lett.} A {\bf 13}, 2771 (1998),\\
C. Schmid, D.J. Schwarz, P. Widerin, {\it Phys. Rev. Lett.} 
{\bf 78}, 791 (1997).
\bibitem{Jedamzik}
K. Jedamzik, J.C Niemeyer, {\it Phys. Rev.} D {\bf 59}, 124014 (1999),\\
K. Jedamzik, {Phys. Rep.} {\bf 307}, 155 (1998), 
{\it Phys. Rev.} D {\bf 55}, 5871(1997).
\bibitem{strangelets}
E. Farhi, R.L. Jaffe, {\it Phys. Rev.} D {\bf 30}, 2379, (1984),\\
E. Witten, {\it Phys. Rev.} D {\bf 30}, 272, (1984),\\
J. Schaffner-Bielich, {\it Nucl. Phys.} A {\bf 639}, 443c (1998).
\bibitem{Annalen} B. K\"ampfer, {\it Ann. Phys.} {\bf 9}, 605 (2000).
\bibitem{Gale} K. Gallmeister, B. K\"ampfer, O.P. Pavlenko, C. Gale,
hep-ph/0010332.
\bibitem{dileptons} K. Gallmeister, B. K\"ampfer, O.P. Pavlenko, 
{\it Phys.\ Lett.} B {\bf 473}, 20 (2000),\\
K. Gallmeister, B. K\"ampfer, O.P. Pavlenko, 
{\it Phys.\ Rev.} C {\bf 62}, 057901 (2000).
\bibitem{CERES} B. Lenkeit (CERES) 
{\it Nucl. Phys.} A {\bf 661}, 23c (1999). 
\bibitem{NA50} E. Scomparin (NA50), 
{\it J. Phys.} G {\bf 25}, 235c (1999),\\
P. Bordalo (NA50), {\it Nucl. Phys.} {\bf A661}, 538c (1999).
\bibitem{WA98} M.M. Aggarwal {\it et al} (WA98), nucl-ex/0006007,
nucl-ex/000608.
\bibitem{Alam} J. Alam, S. Sarkar, T. Hatsuda, T.K. Nayak, B. Sinha,
hep-ph/0008074,\\ 
R. Rapp, E.V. Shuryak,
{\it Phys. Lett.} B {\bf 473}, 13 (2000),\\
R. Rapp, J. Wambach, {\it Eur. Phys. J.} A {\bf 6}, 415 (1999),\\ 
R.A. Schneider, W. Weise, hep-ph/0008083.
\end{thebibliography}
\end{document}